# Prospects of relativistic astrophysics: Viewpoints on the most influential journey

*Dr. Sanjeev Kalita[1,2]*

[1]*Department of Physics, Gauhati University, Guwahati – 781014, Assam*

[2]*e-mail:* [sanjeev@gauhati.ac.in](mailto:sanjeev@gauhati.ac.in)

**Abstract**: Based on several recent developments in relativistic astrophysics and observational cosmology, this article discusses prospects of (1) theoretical structure of General Relativity (GR) in the new era and (2) submission of GR to new and previously unanticipated tests. It explores the internal theoretical challenges experienced by the theory and challenges produced by it in the otherwise successful journey of 100 years. Gravitational collapse, geometrization of laws of physics, absence of strong field astronomical tests, the dark components of the universe (Cold Dark Matter (CDM) + Cosmological Constant ($\Lambda$) or something similar) and cosmological inflation have not only created serious challenges for the theory but also have ignited contemplation for alternatives and attempts for constraining them through observations. The article first explains the nature of the theory and how it was perceived by scientific community a century ago. It then highlights certain perspectives from cosmology and quantum gravity, the ultimate applications of the theory. With the advent of modern observing techniques, the theory has been experiencing tug of war with several tests. Our true journey with the theory has started only recently, thanks to the development of gravitational wave astronomy and extremely large telescopes some of which are and will be dedicated to test the theory by exploring previously uncharted regions of the universe. Incorporating several aspects of observational astronomy and cosmology, gravitational wave research and theoretical physics the article presents the new challenges created by the theory and also their prospects.

# *Contents*

1. **Introduction**

In his seminal article, "Role of General Relativity in Astronomy: Retrospect and Prospect" (1980), Chandrasekhar outlined how the true contents of General Relativity (GR) are hidden in the world of astronomy and how the theory plays inevitable role in the world of compact objects, black holes, gravitational waves and cosmology. Living in the era of rapidly growing field of relativistic astrophysics it will be mere exaggeration to restate the importance of GR in astrophysics and cosmology. However, the theory has been experiencing repeated challenges and also hosted several challenges to astrophysics, cosmology and the fundamental descriptions of gravity. Development of precision observational facilities in astronomy has enabled astronomers to submit the theory to experimental tests in previously uncharted regions of the universe so as to understand whether there is a more encompassing theory broader than GR. The author intends to outline some of these issues in this article.

GR is a theory of the all-pervading force of gravitation which governs our place in the universe. It explains movements of the planets around the Sun, distribution of stars in the sky, formation and evolution of stars in trillions of galaxies and the largest structures of galaxies embedded in infinite space. Einstein's theory, formulated by him in 1915, which is a completion of the old Newtonian theory, has described gravitation as a manifestation of geometry of space and time. Gravitation is a result of mutual interaction of the fabric of four dimensional spacetime and the matter-energy contents of the universe. Matter-energy produces stress in empty space, distorting it. Time, which is a smoothly flowing 'river' in classical physics gets 'compressed' and 'stretched' due to this distortion of space. This distortion of space and time affects the movements of physical objects in the universe. Such a relational philosophy which interrelates apparently disconnected pieces of things was absent in Newtonian world view.

The theory is summarised in the following way. **Space and time participate in the cosmic symphony by responding to the dance of matter-energy, thereby enforcing matter-energy to react accordingly. In the language of John Wheeler, "Matter tells space how to curve and space tells matter how to move."**

A good scientific theory is the one which predicts several new phenomena which are observable and hence making it experimentally refutable. GR possesses this quality. Being a theory of gravity, astronomy has been its natural home (Chandrasekhar, 1980). It has predicted unusual phenomena which are foreign to the Newtonian version and given new pictures of the cosmos. Black holes, gravitational waves, slowing down of time near a gravitational field, bending of light near a massive object in space, origin of the universe in a Big Bang, dark matter and dark energy which are found to shape the galaxies and control the expansion of the universe are exotic predictions of the theory which were not foreseen even by the founder and his followers in the last century. With the advent of modern astronomy equipped with sensitive telescopes both in the ground and space, we have come to know about realities of these ideas.

Recent discovery of gravitational waves – tiny ripples in the fabric of spacetime emitted by merger of black holes has not only established prediction of the theory but also opened a new window to look into previously unexplored regions of the universe. In the coming years we may see a completely new and unexpected picture of the universe, by peering into heart of the black holes or into the earliest moments of the Big Bang.

Even after one century of its successful journey with experimental tests, GR, in its deeper level still possesses enough potential to guide new physical principles for a more complete theory and also to demand new experimental tests. Why has it not been ruled out yet by any observational result? Is it so close to truth?

After discussing the historic challenges experienced by the theory and its mathematical principles, I will proceed little unconventionally (not non-trivially) for highlighting the prospects of relativistic astrophysics.

## 2. Nature of the theory

### 2.1 A history of 'difficulty'

A common lore goes in the following way to appreciate how challenging the theory was since the time of its inception. Ludwig Silberstein, a Polish – American theoretical physicist once asked Sir Arthur Stanley Eddington, the British astronomer and relativist after a lecture delivered by Eddington, " Professor Eddington, you must be one of the three persons in the world who understands general relativity." Eddington was silent for a moment. Silberstein said, "Don't be modest, Eddington." Eddington replied, "On the contrary, I am trying to think who the third person is" (Ferraira, 2014)!

In 1919, through an astronomical observation during a full solar eclipse in South Africa, Eddington confirmed GR by demonstrating that light coming from distant stars grazing over the Sun gets bent due to curvature of space near the Sun as predicted by the theory. In addition the theory also explained the discrepancy of 43 arc seconds in the total periastron advance of Mercury's orbit that could not be worked out by Newtonian gravity. Einstein, however, considered the experimental tests of his theory as peripheral as it was based on elegant physical principles incorporating beautiful mathematics of differential geometry.

 The theory predicted phenomena well ahead of its time. It predicted "frozen stars" – the black holes or compact cold objects formed at the end of stars' lives. It predicted an expanding universe. Lack of observational evidences of these phenomena made it difficult to accept the theory. The telescopes prevalent at that time were not powerful enough to look deep into space and unravel the new phenomena. The theory entered a "dark age". The impression came to the general scientific community that GR is a no more than a playground of abstruse mathematics.

In early 1930s, Indian astrophysicist, Chandrasekhar proposed that stars heavier than a limiting mass should undergo continued gravitational collapse leading to black holes. Eddington, one of the staunchest supporters and leading authority of GR rejected Chandra's idea claiming non-existence of such objects in Nature. In 1935, Chandra presented his new theory of relativistic degeneracy applied to final stage of evolution of Sun like stars in a meeting of the Royal Astronomical Society. He concluded, "A star of large mass cannot pass into white dwarf stage and one is left with speculating on other possibilities." So strong was his belief in nonexistence of gravitational collapse that Eddington dismissed Chandras's idea by saying, "various accidents may intervene to save the star. I think there should be a law of Nature to prevent a star from behaving in this absurd way!". Lack of observation of collapsed objects and impact of Eddington's authoritative work on stellar astrophysics (the standard Eddington model of star, summarised in his classic *The Internal Constitution of the Stars*) delayed the acceptance of Chandra's ideas by about 40 years! Chandra's were foreign to usual expectation of models of stellar structure and evolution.

Around 1939, American theoretical physicist, Robert Oppenheimer and his co-worker, Snyder calculated what would happen to a collapsing star. By applying GR to the late phase of evolution of massive stars, they found that gravitational collapse is inevitable and the final state of collapse is a point of infinite density – a singularity. Both Einstein and Eddington considered it as a beautiful mathematics and abominable physics! No approval from these two *Es* means the theory would be thrown into long grass. Again, due to lack of evidence of collapsing stars till 1960s, GR was considered as foreign to astronomy and was not taught to physics community even in the institution like Princeton.

In 1963, astronomers realised that GR has to be taken seriously. Two important revolutions happened, one in theory and the other in observation. One was the new solution of Einstein's field equations of gravity applied to vacuum (empty regions surrounding a mass point) found by New Zealand's relativist, Roy Kerr and the other was the discovery of quasar.

Kerr's solution generalised Schwarzschild's solution which described spacetime geometry near a singularity formed by non-rotating spherically symmetric collapsing object. The Kerr solution was spinning and twisting and tugging spacetime around it (Kerr 1963). His paper was published by Physical Review Letters which just decided to publish something related to GR only a short time before Kerr's finding! This happened just few months before the first Texas Symposium of relativistic astrophysics that gathered astronomers and relativists together to discuss observational issues related to GR. Kerr presented his solution in a meeting of astrophysicists but was paid little attention, whereas his solution was going to be extremely useful for the astrophysics community. People were so uncomfortable with the final stage of stellar evolution and singularities that solutions of GR equations came out which were singularity free – those of Khalatnikov and Lifshitz. They advocated deviation from spherical symmetry to prevent singularities from occurring. These were, however, turned out to be faulty as proven by detail study of structure of spacetime by Penrose (1965). With his concise and sophisticated mathematical toolkit of differential geometry and topology, Penrose concluded, "Deviations from spherical symmetry cannot prevent spacetime singularities from arising.

The problem was becoming so intense that John Wheeler called it as "the issue of the final state". Chandra, Oppenheimer and Snyder, Schwarzschild and Kerr were all going to be proved correct by astronomers' works in the post 1963 era.

Marteen Schmidt, a radio astronomer observed explosive emission of radio waves emanating from a very compact region of the galaxy Cygnus A. In a Texas meeting of relativists and astronomers, it was settled that radio wave explosion from compact sources in the sky could be explained only if there is a supermassive star undergoing gravitational collapse. This was a twist in the history of GR.

Objects like Cygnus A are called quasars – Quasi Stellar Radio Sources. Quasars were found to be massive and so concentrated that Oppenheimer and Snyder's idea had to be taken into account. They needed a central engine so dense that black hole idea had to be incorporated.

Pulsars were discovered in 1967 as pulsating radio stars whose pulsation period of few to millionth of a second demanded an extremely compact object (the radius $R$ is so small that the gravitational potential, $\phi_{grav} = GM/R$ becomes comparable to square of the velocity of light, $c^2$). These are stars made of extraordinarily compressed (density is of the order of nuclear density $\rho \approx 10^{14} g/cc$) relativistic gas of nuclei which form during gravitational collapse of massive stars as originally suggested by the Oppenheimer and Snyder solution of the problem of spherically symmetric instability in GR. The idea of relativistic instability was taken into account to study the properties of compact stars. Astronomers were convinced with the theory. GR had to be taken into the picture to make sense of newly observed phenomena in the universe. The field of astronomy was enlarged by the annexation of the area of GR. The new field of relativistic astrophysics was born! The field is now augmented by cosmology, extragalactic astrophysics and gravitational wave astronomy.

### *2.2 The maths culture*

Richer the mathematics, simpler becomes a physical theory, larger becomes the number of its predictions of observable phenomena and broader becomes its aspects. Classical mechanics is the marked example. Formulations of canonical transformation, Hamilton – Jacobi theory, action and angle variables, orbital perturbation theory etc. lifted the entire framework of mechanics to the stature of a foundation stone for astronomy and other physical theories! What kind of mathematics has been the guide for GR and what is its prospect?

GR is based on differential geometry of curved manifolds. A manifold, $M$ is a topological Hausdorff space with a homeomorphism, $h : U \to U'$ defined at every point $p \in M$ so that $U \subset M$ and $U' \subset \Re^n$. Here $U$ is the open set defining a local neighbourhood of the point $p$ and the images of the mapping $h$, defined as $h(p)$ has the coordinates $\{x^\mu\}$ with $\mu = 1,2,3........n$. The mapping $h$ is called as chart and $U$ is the domain of the chart. Set of all charts, $\{U_\alpha \,|\, \alpha \in I \}$ constitute an atlas $A$ if $\cup_{\alpha \in I} U_\alpha = M$ (see Chandrasekhar 1992 and Straumann 2009 for details).

Gravity appears if we take a pseudo-Riemannian or non-Euclidean differentiable manifold with local neighbourhood of a point as the Euclidean chart (actually pseudo-Euclidean or Minkowski spacetime). Differentiability of a manifold is a strong criterion to define fields such as the metric or the affine connections. It means that the coordinate changes on an atlas are differentiable ($C^\infty$ type). A richer structure of a differentiable manifold is its diffeomorphism which ensures differentiability of an inverse mapping $\varphi^{-1} : M \to N$ between two manifolds or that of the inverse mapping, $h^{-1}$ between two open sets, $U \subset M$ and $U' \in \Re^n$. Although abstract the maths is, it encompasses the underlying physical principles of GR such as the principle of equivalence (local absence of gravity or local regions of spacetime looking identical to flat Minkowski type). The principle of general covariance (that laws of Nature are same for frames of reference in arbitrary motion with respect to each other) is reflected in the diffeomorphism invariance of the theory.

Manifolds do not take care of any fixed metric space. Here, internal relationship between dynamical entities (such as metric and matter or energy) gives rise to physical processes (see section 4). This makes GR an entirely different theory which is in contrast with non-gravitational laws of physics formulated on a background metric space (the rigid Minkowski spacetime which is unaffected by anything else in the universe), e.g. electrodynamics or laws of fundamental particles (QED or QCD) (Rovelli 1991). It is one of the reasons from mathematics that we are having trouble in unifying (may not be needed as discussed in section 4) quantum theory with gravity. A full quantum theory of gravity has to be formulated in a self-contained differentiable manifold (non-perturbative quantum gravity) and not in a metric space (as advocated by the old perturbative quantum gravity) such that it respects diffeomorphism invariance of GR.

***Mathematics is more of a culture than a discipline***. Irrespective of our fields of research and professions almost all of us have learned set theory and mappings, concepts of limits and preliminary calculus and learned how to draw geometrical patterns such as circles, triangles, lines, curves , rectangles etc. in a two dimensional page. We have demonstrated that sum of the angles of a triangle is $180^0$, however, without asking further why it should be! This is the Euclidean geometry on which the physics of Galileo or Newton is established. It discusses movements of matter in the universe which is described by absolute 3 dimensional flat space and an immutable "time". Space and time in pre-relativity science are absolute backgrounds on which events take place. They affect everything but never get affected by anything. Pre-relativity space and time, therefore, form an inactive rigid substratum.

"Difficulty" is a psychological process to substantial extent. Pre-relativity conception of space, time and matter buried in our mind produces a layer which prevents the mind from perceiving the elegant world as given by GR. We are habituated to maths based on a spacetime whose curvature is small and where particles move very slowly, thus bringing Euclidean geometry into our usual thought processes. These layers are, however, to be discarded as these are only approximations to a higher order but much simpler formalism. Therefore, in new frameworks it is suggested to work in a top-down approach of learning.

Had we started our childhood with non-Euclidean geometry of curved spaces these difficulties could not have occurred. In a top-down approach of learning one can deduce all the laws of physics and geometry from a much simpler and more general non-Euclidean spacetime. Therefore, it is not the maths culture, rather the layers of concepts created by a particular culture which creates difficulties in perceiving the elegant and unified descriptions of Nature.

To start with the elegant or unified description early in our career, a non-trivial psychological change is necessary, probably some kind of "mutation" of the thought processes.

It is true that abstract nature of differential geometry made it difficult to perceive GR as a useful theory early in its journey. But it is also the same mathematics culture which has guided us to look into what should be the nature of fundamental theories.

### *2.3 A good theory is humble*

A good physical theory predicts its own failure. One of the conceptual difficulties with GR is that it has predicted situations where it is barely applicable. Examples are black holes and the Big Bang. **These are boundaries of spacetime with infinite energy density and heat where spacetime geometry has ruptures and perhaps loses its meaning!**

Black holes are the end points of evolution of stars heavier than the Sun. It has infinitely strong gravitational tug on everything of its neighbourhood including light. All things are trapped inside a black hole. On the other hand Big Bang is the primordial point of infinite heat and infinite energy density which underwent colossal expansion heralding the beginning of the universe some 13.8 billion years ago.

The theory becomes undefined near a black hole or the Big Bang. It loses predictive power. As geometry of spacetime loses its meaning we do not have the slightest idea of how matter or energy moves. Time and space have beginning in the Big Bang! No dynamics, no laws of physics!

On the contrary, astronomical observations have revealed that the universe is populated by black holes. Every galaxy including our Milky Way has a supermassive black hole in the centre with mass of a million to billion Suns! Collision between smaller black holes gives rise to burst of gravitational waves which have recently been discovered. When stars are eaten by a black hole, swarm of hot gases of the stars emits energetic radiation as they fall into the black hole. These are X rays or gamma rays which are detected by space satellites such as the Chandra X ray observatory and Fermi Gamma Ray Telescope.

The relic heat of the Big Bang known as the Cosmic Background Radiation has been found to be all pervading in the universe. It has established that the universe emanated from a primordial phase of infinite heat. Tiny ups and downs in temperature of this radiation over the sky, calculated by using the basic framework of the theory have successfully explained the origin of galaxies in the universe.

Failure of GR near a black hole or the Big Bang has given rise to spectacular ideas to predict what lies beyond the beginning of time or what lies inside black hole. One such ambitious program is to provide a granular structure of space and time. Just as any matter is granular, being made of discrete atoms, space and time are assumed to have "atomic" structures in scales trillion times smaller than a normal atom. In scientific community it is known as the quantum spacetime. Quantum spacetime ("particles of space" or "particles of time") is not yet matured enough to become a testable physical theory.

Why does spacetime look like the one described by GR in the macroscopic universe? This question is going to push the theory beyond the horizon, to the next level of richness in the coming century.

Astronomers have yet to test the theory very near to a black hole. That is going to be the observing program of major astronomy projects of near future. Post 2020 era will see operation of extra-large telescope (with mirror size of 30 metre or more) which will be dedicated to test the theory near black holes or in very large scales of the universe. It will perhaps answer why the theory has been so successful or see if there is any deviation. We are beginning our true journey with GR only now!

## 3. Cosmological perspectives

### 3.1 Not having a theory of the universe!

We still do not have a theory for the beginning of the universe. Big Bang is only a successful model, parameters of which cannot be predicted as the theory fails in the initial moments of the universe. These quantities are inferred only from astronomical observations! If the theory is a failure during the universe's birth, how does it give a correct picture in the present time, after 13.8 billion years of the Big Bang? The tension is the following. Several independent predictions of the theory for a given measurable quantity of the universe give rise to congruent results! If two uncorrelated experiments for one quantity give rise to similar values of the quantity, the theory with which the experimental data is analysed must be taken seriously. We still do not have a consistent answer to this aspect of the theory. One example is as follows.

Cosmic expansion rate, expressed by the Hubble parameter, $H_0$ is a model independent quantity, simply governed by homogeneity and isotropy of large scale spacetime (the FLRW metric) whose expansion is described by only one parameter, the scale factor $a(t)$. It is determined by the amount of matter content of the universe, the density parameter, $\Omega_m$. The latter is inferred from the luminosity function of the galaxies, dynamics of large scale structures and fraction of hot gases in rich clusters of galaxies (Peebles 1993; Weinberg 2008). Both of these parameters individually do not assume any theory of gravity to get inferred. But their relation does. The way expansion rate is determined by matter density of the universe depends on a theory of gravity – the field equations. The relation is given by the standard Friedmann model of the expanding universe. Observational data have not yet contradicted the connection given by GR.

Cosmology, however, requires extreme carefulness while drawing inferences on the history of the universe. The Friedmann model works well from around 1 sec after the Big Bang. This is the epoch when the universe was dominated by protons, neutrons, electrons, neutrinos and photon, which are fairly understood stuffs in micro-physics. This is the time of formation of the light nuclei in the primordial nucleosynthesis process which predicts the relative abundances of the pristine elements - deuterium, helium and lithium. Observations on the metal poor stars and quasar spectra have given strong empirical support to Big Bang Nucleosynthesis, thereby establishing the model.

However, the Friedmann universe does not make any testable prediction for the ultra-early universe when its age was smaller than fraction of $10^{-3}\,\sec$. The energy scale of the universe at such early times is so high that the micro-physics of fundamental particles that could have survived at that time has not yet been fully understood. If we extrapolate the model far back in time we enter the domain of speculative science of quantum gravity where the Hubble length $cH^{-1}$ becomes comparable to the Planck length, $l_P = \sqrt{G\hbar/c^3} \approx 10^{-33}\,cm$. Putting constraints on the physics of such early time is far beyond the capability of the present terrestrial accelerators (Rees 1995).

### 3.2 The dark universe of GR

If GR is correct then 96% of the universe is completely "dark"! Observations of movements of stars in galaxies or movements of individual galaxies in clusters of galaxies have shown that these movements are too fast to be tolerated by gravitational grip of luminous matter. In the framework of GR it inevitably demands existence of completely different types of matter whose gravity holds on the systems from disintegrating. Known as "dark matter" it is found to be nothing like atoms that we are familiar with. Till now we do not have a proper theory of "dark matter" except the fact that they have to be non-baryonic. Neither we know what their particle contents are nor do we have the idea of their distribution in the galaxies.

Gravitational lensing can provide us with detail map of dark matter distribution in the scales of galaxies and their clusters. Galaxies and massive clusters act as lenses which distorts the images of the background quasars and galaxies by strong and weak lensing. Assuming GR as a correct theory in these scales, lensing data can be used to chart the gravitational potential caused by the lenses and thereby identify the correct distribution of dark matter. Upcoming observing facilities such as the Thirty Meter Telescope (TMT) and Large Synoptic Survey Telescope (LSST) will be dedicated to constrain the mass profile of dark matter in galaxies through more number of gravitational lenses (Skidmore 2015). If there is modification to gravity, it will show up in additional gravitational degrees of freedom which are possibly locked in subgalactic scales by some screening mechanism. These may be non-universally coupled fields (opposed to gravity) which cause a deviation of mass of a structure determined from dynamics of subsystems from the mass determined from lensing. The dynamical mass is given by $M_D \approx \langle V^2 \rangle R / G$  where, $V$ is the average velocity of the subsystems (random velocity of stars in galaxies or galaxies in a cluster) and $R$ is the size of the structure. The lensing mass, $M_L$ depends only on the bending angle of light, $\alpha$ as observed by the distant observer which is expressed as $\alpha = GM_L / c^2 f(d)$ where, $f(d)$ is some function of distances between lens and the source, source and the observer and observer and the lens. Till now we have found no difference between lensing and dynamical mass. However, improvements in astrometric (position measurement accuracy during lensing) and Doppler accuracy (velocity estimation) of the large telescopes will be able to detect tiny difference (if any). Whereas null results will restore GR, positive one will demand re-examination of the dark matter interpretation.

Another mystery confronts modern cosmology. Estimations of curvature (measurements of deceleration parameter, $q_0$) in relativistic redshift- magnitude relation (m – z) of Type Ia supernovae (Riess et al. 1998; Perlmutter et al. 1999)  have shown that the cosmic expansion is accelerating. It has dethroned the long held expectation that cosmic expansion would slow down due to gravity of "dark matter" in the galaxies. The runaway cosmic expansion has called for an energy component which acts against gravitation – **a cosmic repulsion**! If GR is the correct theory of gravity this must be due to either energy in empty space – the vacuum energy or some exotic form of energy having negative pressure. This is commonly known as "dark energy". We do not have the knowledge of why it should be. The energy in the vacuum is calculated by using the quantum theory of particles and is found not to look like anything inferred from the astronomically measured value. They differ by 120 orders of magnitude – the greatest hierarchy between theory and observation that has ever been realised!

In GR, empty space is modelled by the "cosmological constant" ($\Lambda$) which provides the fabric of spacetime with a zero point curvature such that space attains an intrinsic tendency to expand exponentially. Therefore, the "cosmological constant" acts as a geometric candidate for accelerating the universe. The problem lies with the interpretations of $\Lambda$. First, it sits as a curvature term in the left hand side of Einstein's field equations. Second, the moment we transfer it to the right hand side containing sources (matter fields), it gravitates as "matter-energy density" which is a component of the energy- momentum tensor. It is the second aspect which motivates theorists to calculate it as a vacuum energy of quantum fields. However, it gives rise to an energy density 120 orders of magnitudes larger than the observed value. In the first aspect, we do not have to ask where from $\Lambda$ comes and what its magnitude is, as it stands as a constant term similar to the Cavendish constant ($G$). But it calls for a deep question – where does the classical geometry come from? For both of them we do not have a satisfactory answer.

Is it the signal that GR has some crack? We are not yet sure. All observational data make sense if we include "dark matter" and "dark energy" in our equations. However, we do not have a complete physical theory of these dark components. They constitute 96% of the energy budget of the universe. The "dark universe" is, therefore, another complexity that has arisen in the theory.

It might be a possibility that we are using a wrong background spacetime in interpreting cosmological data. The maximally symmetric Friedmann – Lemaitre- Robertson – Walker spacetime is the background of $\Lambda CDM$ Big Bang cosmology. It also constitutes the background around which density perturbations needed for galaxy formation, are calculated. But inhomogeneous models such as the Lemaitre – Tolman – Bondi spacetime can explain the cosmic acceleration without invoking dark energy. These issues will be settled to high degree of accuracy by future observations.

The problem of the dark universe has been a focal point of upcoming extremely large ground based telescopes including TMT, LSST and the new generation radio facility in the form of Square Kilometre Array (SKA). It is going to survive as a major problem in astronomy and cosmology, at least for coming few decades!

### 3.3 Is it too much of faith?

There is a lesson from history. Attempts to understand the orbital anomaly of Mercury in 19[th] century within the framework of Newtonian theory of gravity gave rise to a temporary belief of existence of Vulcan – a hypothetical planet between Mercury and the Sun. Celestial dynamics was used to its fullest power equipped with orbital perturbation theory to explain Mercury's anomaly. However, Vulcan was ruled out by a new theory of gravity. Are we being misled in the same way? **It is not irrelevant to form a class of thoughts to consider that dark matter and dark energy simply evaporate in a new theory of gravity beyond GR. Gravity and spacetime itself can manifest in a new way in the relevant cosmic scales which has been interpreted as due to dark stuff.** Theoretical attempts to bypass the dark stuff are of many forms. Examples are theories of gravity with several gravitational degrees of freedom (scalar – tensor gravity and geometric extensions to Einstein gravity). Very recent and unconventional attempt is the one reported by Maeder (2017). In this approach the dark stuff can be mimicked by scale invariance of empty space – the property of space which remains unaltered during expansion and contraction. This model has been found to reproduce the large scale motions of galaxies without resorting to the dark stuff of matter and energy (Maeder 2017).

It is very true that we should not always extrapolate history forward. After all we have at least Neptune! Even if existence of dark matter and dark energy are considered to be firmly established we do not yet know how to accommodate them fundamentally in the theory of gravity we are accustomed to. For example, it is still to be experimentally or observationally tested whether dark matter respects the principle of equivalence (universality of free fall) on which the theory is based (Perez, Doser and Bertsche 2017). Within the framework of GR, dark energy, in addition to the $\Lambda$ term, is well modelled by dynamical fields (quitessence, quintom, phantom, chameleon etc.) with time varying equation of state ($\omega(t) = p(t)/\rho(t)c^2$). We are waiting for observational data of upcoming facilities such as Dark Energy Survey (DES) and LSST to confirm whether dark energy is dynamical. If yes, the next level of discussion will be whether these fields are universally coupled to all forms of matter-energy as demanded by the principle of equivalence (universality of gravity).

That gravity itself can behave as dark matter or dark energy is a new idea which is being exercised both in theory and observations. We have a recent case where galaxy – galaxy gravitational lensing observations in KiDS and GAMA survey (Brouwer et al. 2016) have explained the data with Emergent Gravity (EG) hypothesis of Verlinde (2016) . According to this alternative of GR, gravity and spacetime are macroscopic (effective) notions which evaporate in microscopic scales. The EG hypothesis yields excess gravity in the scale of galaxies and clusters of galaxies without resorting to dark matter. Although dark matter idea fits to the data well when compared to the EG hypothesis, the later does so with smaller number of free parameters than needed by the dark matter hypothesis. Of course, it is not yet promising to rule out dark matter from the cosmological scenario. But it opens the door to future cosmological observations for putting microscopic description of gravity to test.

Any faith will, therefore, be judged by data!

### 3.4 Inflation and galaxy formation

Another situation where, I feel, GR has been extrapolated with too much of confidence is the inflationary expansion phase of the primordial universe. Whereas inflation predicts large scale homogeneity and flatness of the universe as demanded by observations, the very mechanism of a superluminal expansion around $10^{-34}$ sec of the Big Bang, which erases primordial curvature and irregularities, requires quantum phenomena. Inflationary expansion is found to be generated by scalar fields (known as *inflaton*) which are thought to have popped out of existence due to quantum fluctuations of the vacuum. One central role inflation has played is its prediction, in addition to homogeneity and flatness, of a scale invariant density perturbation (Harrison – Zeldovich) spectrum which seeded the growth of gravitational instability leading to the formation of galaxies in the late phase. The metric perturbation $\psi = \delta\phi_{grav}/c^2 \approx 10^{-5}$ in the large scale universe appears naturally under some choice of the *inflaton* potentials. But it is philosophically unappealing to apply a classical theory to dynamics of quantum fields! The inflationary expansion is achieved by using classical Friedmann models to quantum fields which act as sources in the field equations of gravity. In a sense, inflation is a semi-classical model invoked to explain the large scale structure of the universe. We do not have a full theory of gravity or matter fields from which inflation is derived. Perhaps this is the reason inflation requires parameter adjustment to generate right amplitude of initial perturbations that can produce the final shape of matter power spectrum of galaxies as observed in large scale surveys. It is expected that detection of primordial gravitational waves generated during inflation will be able to judge the exactness of GR in the primordial universe.

### 3.5  Cold Dark Matter (CDM) or gravity?

The current scenario of galaxy formation resorts to non-relativistic, nonbaryonic Weakly Interacting Massive Particles (WIMP) as candidates of CDM. It has given rise to the hierarchical cosmogony where galaxies or subgalactic structures (massive black holes similar to those existing at the galaxy cores or the globular clusters) are the first objects to form in the universe through condensation and dissipation in CDM halos. Existence of high redshift quasars (see Banados et al. (2017) for discovery of the highest redshift supermassive black hole) has established the CDM hypothesis. Right amount of CDM abundance ($\Omega_{CDM} \approx 0.28$) in computer N – body simulation has been able to generate the observed matter power spectrum of the universe. However, the resolution of the present simulations is poor enough to probe the scales relevant to galaxy cores or the massive black holes that formed in the early universe.

Time to time we have seen some antitheses to CDM paradigm. Absence of large number of satellite (dwarf) galaxies near massive galaxies (Kylpin et al. 1999; Moore et al. 1999; Bullock 2010), wobbling of Bright Cluster Galaxies (BCGs) (Harvey et al. 2017) are few marked examples. Irrespective of these possible observational challenges, CDM can be replaced by Self Interacting Dark Matter (SIDM) which can form 'dark matter cloud' around the galaxy cores harbouring massive black holes ( Saxton, Younsi and Wu 2016). The best candidate for self-interacting entity is gravity itself! If gravity has departure near black hole singularities, it naturally calls for additional degrees of freedom, e.g. scalaron fields ($\psi$) in $f(R)$ theories. Under the condition that the Compton wavelengths of the fields are comparable or larger than the gravitational radius of the black hole, they can form stable bound clouds near the hole (East and Pretorius 2017). It has been recently reported (Kalita 2017 a) that scalarons with mass $M_\psi \approx 10^{-22} - 10^{-20} eV$ can mimic dark matter cloud which perturbs light bending near the hole that could be detectable through the astrometric observations of the upcoming large optical/IR ground based telescopes. ***Thus if dark matter is an anomaly with gravity, not exotic particles, we must look into a larger framework of gravity beyond GR.***

## 4. The problem of unification

### *4.1 "Marble" to "wood" or wood" to "marble" ?*

In his later years, Einstein tried to go beyond GR for a unified theory of all the forces of Nature. His motivation was to replace the quantum theory (Smolin 2015) which advocates that microscopic laws of particles constituting matter are based on uncertainty where cause and effect relations of physical events dissolve. Einstein was quite disappointed at this point. However, his geometrical path to forces of Nature was proved to be a failure.

Einstein's field equations of gravity connect matter-energy of the universe to geometry of spacetime. The equation is written as a relation between spacetime curvature described by Riemannian geometry and the energy density and pressure of matter distribution in the universe:

$$R_{\mu\nu} - \frac{R}{2}g_{\mu\nu} + \Lambda g_{\mu\nu} = \frac{8\pi G}{c^4}T_{\mu\nu} \qquad (1)$$

where the left hand side is the geometry expressed by the spacetime metric $g_{\mu\nu}$ and the quantities derived by its derivatives (the Ricci tensor, $R_{\mu\nu}$ and the scalar $R = g^{\mu\nu}R_{\mu\nu}$). The third term in the left hand side contains the cosmological constant $\Lambda$ which causes additional twist in the curvature of empty space which is one of the challenges not only for the theory but also for modern cosmology. The tensor $T_{\mu\nu}$ in the right hand side is the energy-matter tensor containing pressure, energy density and momentum flux as its components which constitute the very cause of curvature and dynamics of spacetime.

The right hand side is "Wood" made of discrete particles whereas the left hand side is "Marble" – the geometrical structure of spacetime. Here lies the deep mystery of the theory. **We are equating things having different internal patterns**! One approach is to give a quantum description to geometry- conversion of "Marble" into "Wood". This is the process of quantisation of geometry. The other approach is the one of Einstein – to give a geometrical pattern to the fundamental forms of matter. This is the process of converting "Wood" into "Marble". Neither quantisation of geometry nor geometrisation of matter has been possible yet. This piece of science will continue until we have a unique pattern of both the things.

### 4.2  Geometrisation

The basic reason behind our respect to geometrisation is that by this process, apparently uncorrelated phenomena get unified into a single whole. In history we have the lesson that adding time as an additional dimension to space unifies electric and magnetic fields into a unified phenomenon of electromagnetism. Maxwell's theory is thus a by-product of increasing the dimension of the universe from 3 to 4. Attempt to geometrise electromagnetic fields involves one more spatial dimension as originally proposed by Kaluza and later extended by Klein into a 5 dimensional world in the so called Kaluza–Klein (KK) theory. In KK theory Maxwell's electromagnetism appears naturally as a coordinate transformation (not as U(1) group of gauge transformation of fields in ordinary electromagnetic theory) in 5 dimensional spacetime. The photon arises from the 5 dimensional metric tensor, $g_{AB}$ ($A, B = 0 - 4$). This is the geometrisation of the gauge theory of electromagnetic fields.

The KK theory is a minimal extension of GR to incorporate the photon. Development in theoretical physics in post 1970s saw richer extensions of GR which not only unifies other gauge forces (present in SU(3) and SU(2) non-Abelian gauge theories) with gravity but also relates fermions to bosons and vice versa. These are supersymmetric extensions known as supergravity theories. These ideas require seven extra dimensions of space and supersymmetric particles in the universe. No one has the slightest idea of why the universe should have 11 dimensions. But the physics of the forces and matter in the universe becomes simple with this number.

This is one of the approaches for quantum gravity. There are several approaches to it, however, without having a convergent result. There is neither any signature of extra dimensions of space nor of supersymmetry in laboratory experiments.

But it seems certain from differential geometry consideration that gravity has to be accommodated in a higher dimensional manifold, that of at least 5 (Dadhich 2005). The reason is the following. The Riemann tensor which describes the curvature of space and hence the gravitational field requires minimum 2 dimensions to survive. But the matter fields already occupy the 3 dimensions of space, leaving time alone. Thus one extra dimension of space is required for a full theory. Therefore, GR has to be extended to provide a more general description of gravity.

### 4.3 Can beauty be a guiding principle?

GR hosts conceptual challenges in the process of unification. The theory was formulated by Einstein by basing it in well-defined physical principles – the principle of general covariance and the principle of equivalence which require no sophisticated mathematics, rather some "thought experiments". These are, however, experimentally falsifiable principles. However, the journey of unification of GR with quantum theory has been pursued till now with 'no physical principle' (except the relatively recent realisation of *holographic principle* and the *principle of background independence* (Smolin 2015 and references therein)). It is being continued with an expectation that as we probe deeper into spacetime and matter more and more symmetries will uncover. For example, $SU(3) \times SU(2) \times U(1)$ of standard particle theory going to SO (32), $E8 \times E8$ or opening up of exceptional infinite dimensional duality symmetry group, E10 which makes sense even near the Big Bang type singularities, are few examples which are being extensively exercised in the journey of unification. However, we have failed to find these monstrous symmetry groups. In Nature, they might have broken spontaneously which requires adjustment or fine tuning of parameters such as particle masses, coupling constants, vacuum energy etc.

Why to pursue, then, something called unification? The motivation lies in history that the underlying principle of symmetry has been instrumental in formulating Maxwell's theory (Abelian), Yang- Mills theories (non-Abelian) and even GR itself! This is, however, only an aesthetic guide and not a physical principle such as the covariance and equivalence.

Too much of beauty (a form of *maya* in eastern philosophy) can spoil or blur our perception of truth. We should not forget that Einstein's failure in unifying forces is associated with his journey with pure mathematical principles! The same Einstein excelled in his special and general relativity due to his grip on physical principles. What is the physical principle (or is there any?) of unification that we are striving for? It is difficult to present any. Smolin (2005) put forward the idea of *background independence* as a physical principle behind quantum gravity in the same line of principle of general covariance of GR. It says that laws of Nature must not rely on fixed background geometry of spacetime. Clearly GR belongs to a *background independent* theory. It is the intrinsic geometry (the metric) of four dimensional curved manifolds which manifests as gravity and governs motion of matter fields. It is the harmonious dance of metric and matter fields. The background of Newtonian physics (absolute space and absolute time) or special relativity dissolves into a dynamical metric field. The metric and matter form a dynamical pair. They do not merge!

Therefore, closer to the truth, one may expect lesser unification rather dynamical relationships (Leibnitz's relational philosophy). ***No unification, only relations***! Absolute beauty will then spread out into shared beauties. This idea manifests in the Loop Quantum Gravity (LQG) (Rovelli 2010; Smolin 2000) where network of "atoms" of space, loops around them and their interrelations give rise to quantum description of geometry. In this approach there is no unification of forces rather only geometrical relation is the starting point.

LQG thus respects the original philosophy of *background independence* of GR. This is another philosophically challenging journey GR has initiated.

### *4.4 Stars and quantum gravity !*

One of the remarkable problems in relativistic astrophysics is the natural emergence of upper mass limit of cold, degenerate and compact stars. The structure of such objects is quantified by relativistic hydrostatic equilibrium (Tolman – Oppenheimer – Volkoff equation). This is the equilibrium between gravity and gas pressure described by the polytropic equation of state, $P = K\rho^{1+1/n}$ where $K > 0$ and is characteristic of the gas. It is known as the poytropic constant and $n$ is the polytropic index specifying the stellar model. Condition for stability of these objects demands an upper mass limit which comes out as $M^* = kM_{Pl}$ , where $M_{Pl} = (\hbar c/G)^{1/2}$ is the Planck mass and $k$ is dimensionless parameter related to the specific gas properties (mean molecular weight, particle masses etc.) and hence determined by the polytropic constant. It means that for $M > M^*$ there exists no positive pressure solution for counteracting gravity, thereby leading to continued gravitational collapse. It is quite marvellous that the three fundamental constants $\hbar$ , $c$ and $G$ naturally occur in the same equation for the upper mass limit. The first one appears due to quantum statistics of the degenerate gases (electrons for white dwarfs and neutrons for neutron stars) and the other two appear due to the relativistic theory of gravity. Their simultaneous occurrence is observed nowhere except in quantum gravity where the gravitational length, $GM/c^2$ becomes comparable to the Compton wavelength, $\hbar/Mc$ . Therefore, the connection between upper mass limit of compact stars and the Planck mass is non-trivial and is related to deep questions in cosmology (see Kalita 2017 b).

## 5.  Spacetime and astronomy

Let me come to the prospects now. Whether it is cosmic inflation, generation of initial perturbations, modified gravity and loops of quantum gravity, our first challenge will be to have strong field astronomical tests of the theory (even ultra-strong field, $\phi_{grav} = c^2$ lurking near the event horizon of black holes) to see the nature of the more general theory of matter and spacetime. Some of the current frontiers are discussed below.

### *5.1 The Galactic Centre as a laboratory*

The simplest spacetime of GR is the vacuum solution of the field equations (with $T_{\mu\nu} = 0 = R$, assuming a zero cosmological constant) which are the black holes. Discovery of a massive black hole of around $M = 4 \times 10^6$ M$_\odot$ at the Galactic Centre (Sgr A*) by infrared telescopes (Ghez et al. 2008; Meyer et al. 2012) has given astronomers a unique laboratory to test GR and its different metric forms. All astrophysical black holes are described by the Kerr solution of GR – black holes characterised only by two parameters, mass (M) and angular momentum (J). It is known as the "no – hair" theorem or the uniqueness theorem which asserts that vacuum solutions of all metric theories are described by the Kerr solution. However, it is still not clear from observation whether the black holes are described by Kerr spacetime. The "no-hair" theorem is yet to be strongly established by observation. Independent measurements on spin of the black holes are of fundamental importance for testing the theorem and a deep understanding of space and time. Upcoming ground based observatories such as GRAVITY in Very Large Telescope (VLT), TMT and European Extremely Large Telescope (E- ELT) possess sufficient potential for measuring spin of the massive black hole, Sgr A* (Will 2008; Yu, Zhang and Lu 2016) through astrometric observations of the relativistic stellar movements (Doppler shift and precession of the periapses) near the Galactic Centre. It is briefly discussed below.

According to the "no – hair" theorem, spin ($J$) and quadrupole ($Q_2$) of the black hole are related as $Q_2 = -J^2 / M$ (in the unit of $G = c = 1$). Thus accurate measurement of the contribution of spin and quadrupole to the motion of test particles near the black hole is required for testing the theorem. Both spin and quadrupole modify the pericenter advance of masses traversing Keplerian orbits around the black hole. With advent of adaptive optics in ground based facilities such as the Keck and VLT, astronomers have been able to resolve a few stars orbiting the central black hole of our Galaxy. These stars act as test particles to probe the spacetime near the hole. Due to rotation of the black hole (unknown) the stellar orbits experience additional pericenter shift due to spin (Lens-Thirring or the frame –dragging effect) and quadrupole. These are added to the normal pericenter shift that can be calculated from the Schwarzschild geometry (non-rotating black hole). The astrometric sizes of these effects are given by the following periastron shift rates,

$$\dot{\theta}_{prec}(Sch) \propto a^{-1}(1-e^2)^{-1} \qquad (2)$$

$$\dot{\theta}_{prec}(J) \propto Ja^{-3/2}(1-e^2)^{-3/2} \qquad (3)$$

$$\dot{\theta}_{prec}(Q_2) \propto J^2 a^{-2}(1-e^2)^{-2} \qquad (4)$$

where, "*Sch*" is for the Schwarzschild effect and $a$ is the semi major axis of the Keplerian orbit. It is clear that accuracy in the measurement of spin and quadrupole effects requires tracking of high eccentricity and short period (small $a$) orbits. The next generation of optical/IR instruments such as GRAVITY in VLT and TMT will surpass the existing resolution and astrometric accuracy. They will be able to resolve stars within few *mas* of the Galactic Centre. It has been found (Will 2008) that extracting spin and quadrupolar gravity requires astrometric accuracy of about $10 \mu as / yr$ for the precession effect. Spin also affects the lensing of star light during a star-black hole occultation. For a star at 50 AU orbit (lens –source distance) and for the range of dimensionless spin parameter, $\chi = J / M^2 = 0.1 - 2.0$ (from sub Kerr to extreme and super Kerr), the position shift of the star on the sky plane due to lensing falls in the range 21- 22 $\mu as$ (taking source distance as the distance to the Galactic Centre, $D \approx 8 Kpc$ ). Astrometric potentials of TMT and GRAVITY will be able to detect such effects and hence test GR for the first time near the most exotic object in the universe. It has been recently demonstrated by this author (Kalita 2017 a) that even a modification to gravity with additional gravitational degrees of freedom will be testable by such instruments if the additional degrees of freedom are described by scalar modes with mass, $M \approx 10^{-22} - 10^{-20} eV$ . These are modes present in geometric modification to GR such as the $f(R)$ theories (Starobinsky 2007; Gannouji, Sami and Thongkool 2012). These theories require existence of black hole hair (scalar hair more specifically) which contradicts the "no-hair" theorem. Therefore, observation will justify whether the "no-hair" theorem is perfect or the *hairs* are screened at high curvature regimes!

Our true observational journey with GR has started only recently. We have now proper astrophysical laboratory to probing spacetime deeper – in regions which possibly host departure from the conventional ideas of gravity.

### 5.2 Gravitational collapse: Is it solved?

Since 1963, gravitational collapse has been recognised as a central problem in relativistic astrophysics. Vacuum solutions of Einstein's field equations which are stationary, are uniquely described by the Kerr solution. These solutions harbour a central singularity cloaked by a horizon – known as the "cosmic censorship hypothesis". It is however, still a matter of intense debate whether all black holes are of Kerr type or the universe may have "naked singularities", those without horizon! Settling of this issue requires accurate measurement of the black hole spin ($\chi$). Existence of Kerr horizon requires that black hole spin can attain the maximum value of $\chi = Jc / GM^2 = 1$ (in proper units). It is called as extreme Kerr solution. Observation of extreme Kerr black holes, therefore, opens the possibility of not only naked singularities but also of their role in astrophysical processes.

Bardeen (1970) worked out that efficient accretion of matter onto black holes at galaxy centres may form such extreme Kerr objects. This proposal underlies the idea that extreme Kerr black holes may be precursor of energetic jets of AGNs (Benson and Babul 2009). X Ray satellites such as the Chandra X Ray Observatory and NuSTAR have detected accretion disks of hot gases circulating around possible black hole candidates in our Galaxy. Gases undergoing accretion give a unique opportunity to measure the spin of the central black hole. The idea is that for rotating black hole, location of the innermost stable orbits of the accreted gases depends on the spin. X rays climbing up from the innermost edge of the accretion disk and hence undergoing gravitational time dilation, therefore, carry information of how fast the hole is rotating. The effect is, however, masked by presence of hot gases in the exterior of the disk with which the X ray photons undergo energy loss.

With the capability to disentangle these two effects, NuSTAR has measured spin of the supermassive black hole at the centre of the galaxy, NGC 1365 with the lower limit $\chi \geq 0.84$ (Risaliti 2013). Relativistic Magneto Hydrodynamic Simulations show that galactic black hole spin can be as large as $\chi = 0.94$ to cause relativistic jets (see Benson and Babul 2009 and references therein). How does relativistic spin grow? Formation of supermassive black holes in the universe is still a challenging problem for the cosmogonic scenarios. But a rapidly spinning black hole indicates a violent phase of accretion. Therefore, prospect of spin measurements is not only of understanding the nature of spacetime near the hole but also of unravelling the formation and co-evolution of black hole – galaxy systems and active galaxies as well.

### 5.3 Gravitational Wave Astronomy

Detections of gravitational waves from four binary black holes and one neutron star merger event are important findings for frontiers in astronomy. Discovery of gravitational waves, however, is not a test of GR. Rather we are taking GR as a background model to analyse the wave forms. What is new then? It is not a new avenue for GR rather a new window to do astronomy by learning about exotic sources. That is what gravitational wave astronomy is.

One of the theoretical challenges in gravitational wave research is to ensure that the linear spacetime metric perturbations hold good in the full non-linear field equations of GR. Although existence of waves in the full theory has been shown to exist long time back (Bondi, Pirani and Robinson 1959; Pirani 1957; Robinson and Trautman 1960), it requires challenging numerical computation. Result is the development of numerical relativity as a new branch of relativistic astrophysics.

The gravitational waves emitted by coalescing black holes carry signature of three phases – inspiral, merger and the ringdown. Linear calculations are accurate at the inspiral phase but needs to be significantly improved as the velocity increases. Higher order terms in the Post- Newtonian expansion (expansion of the gravitational wave amplitude in terms of power of V/c) of the metric appear and they give rise to corrected waveforms. The merger phase is tedious to calculate as V~ c. Computationally intensive numerical methods have to be invoked to study this phase. GR based wave forms generated can be used to study the sources involved – binary black holes, black hole – neutron star systems and binary neutron stars, thereby leading to new avenues in astrophysics.

Each of the three phases gives idea about the sources. However, numerical relativity calculations such as those performed in Simulating eXtreme Spacetime collaboration consumes weeks or even months with 20,000 – 70,000 CPU hours (Buonanno 2017) to compute waveforms. While more theoretical work is needed to probe deeper into the sources, experiments such as those in aLIGO, VIRGO and upcoming gravitational wave observatories will also need to improve accuracy to test gravity itself in the strong field regimes.

This is a new journey initiated by the theory and its consequences for astrophysical objects. GR, however, has to be resubmitted to precision experiments to see whether there is post-Einstein departure. It is because we are using the standard theory as the background model to analyse the signals, whereas it may have more encompassing extensions close to the black hole singularities.

### 5.4 Gravitational waves and dark stars

The bottom – up scenario of structure formation (hierarchical cosmogony) requires condensation of subgalactic sized dark matter halos ( M~$10^5$ – $10^6$ M$_\odot$) in the early universe. Formation of galaxies then proceeds through merger of these systems. These are the primordial massive black holes. Irrespective of what dark matter is, existence of massive black holes with M~ $10^8$M$_\odot$ in the early universe (Banados et al. 2017) at least confirms the bottom – up approach. Merger of these black holes is associated with gravitational wave emission at extremely low frequency, $f \approx 10^{-3} - 10^{-4} Hz$ which will perhaps eventually be discovered by space based interferometers such as LISA. But the challenge will be the event rate. It is not yet clear how the primordial massive black holes formed.

One radically new idea is that these black holes grew from seed intermediate mass black holes with M ~ $10^3$ – $10^4$ M$_\odot$. These are recently proposed to be result of direct gravitational collapse of extremely massive first generation stars made purely by WIMP dark matter and pristine hydrogen gas (Freese et al. 2015). Called as dark stars these are shown to be capable of growing in mass up to 100,000 M$_\odot$ unlike the fusion guided stars. If this is the case, the growth of massive holes has a basis. Future gravitational wave astronomy will, thus, have the challenge of detecting binary intermediate mass black holes in the early universe.

### 5.5 Is GR breaking down?

Recent analyses of publicly available data from LIGO detector have shown (Abedi, Dykaar and Afshordi 2016) that there may be deviation from GR at the scale of event horizon (the gravitational radius of the hole, $R_g = 2GMc^{-2}$). In GR, the event horizon is a stable region surrounding the singularity, defined by the null rays, which separates two regions of spacetime – the one from where no physical information can escape to the exterior world and other being the exterior region from where something can fall into the hole. In some versions of quantum gravity, however, the event horizon is constituted by microscopic degrees of freedom (strings in "string theory", loops in LQG etc.) and display a fuzzy structure made by fluctuating high energy particles. It is known as "firewall". Existence of a "firewall" affects the propagation of gravitational waves. This causes trapping of the wave signals and re-emission after the initial burst of the waves. Result is the formation of "echoes" in the waveforms. These "echoes" have been reported in the recent analyses (Abedi, Dykaar and Afshordi 2016), which may be statistical fluke but could be a matter of serious concern for the theory if they survive in future detections of waves from black holes.

This observation indicates possibility of detection of departure from GR in future gravitational wave observations.

### 5.6 Is the final farewell to $\Lambda$ in sight?

In the history of GR and cosmology we see several occasions of withdrawal and resurrection of the $\Lambda$ term. Einstein's withdrawal of the cosmological term from the field equations (1917) was followed by inclusion in the classic loitering cosmological model of Eddington and Lemaitre. It helped in explaining overabundance of quasars at cosmological redshift of around $z = 2$. It was then again forgotten due to significantly successful Einstein-de Sitter cosmology which accounted for almost all observational data in cosmology up to 1990s except the cosmic age crisis and the galaxy distribution. Astronomers reconsidered $\Lambda$ once again to resolve the age problem and explain the large scale structure. It surprisingly worked! A $\Lambda$ dominated universe appears much older than the oldest stars (see Amendola and Tsujikawa 2010 and references therein). It was needed in N- body computer simulations (Efstathiou, Sutherland and Maddox 1990) with an energy density parameter of around $\Omega_\Lambda \approx 0.8$ to generate the right distribution of galaxies in large scales. Then came the historic discovery (for which one needs not any introduction) of Riess et al. (1998) and Perlmutter et al. (1999) that it is difficult to explain the recent phase of cosmic acceleration without the $\Lambda$ term. ***It was a good beginning which symbolised that something big would be happening in cosmology.***

Irrespective of these ups and downs, the cosmological constant survives in the very fabric of spacetime as a zero point curvature ( $R_{\mu\nu} = \Lambda g_{\mu\nu}$ for vacuum with $T_{\mu\nu} = 0$ ). If a fundamental theory demands dissolution of spacetime is it really needed or we have already signature of some alternatives to interpret data? Although it is too early to connect them to the new theory, I will highlight two observations which tells about possibility of ruling out $\Lambda$ .

First, measurements of the Baryon Acoustic Oscillation (BAO) (acoustic perturbations in the primordial baryon-photon fluid which was frozen due to removal of the radiation drag at decoupling) at various cosmic epochs by the SDSS III (BOSS) survey (Zhao et al. 2017) have shown evidence of time varying equation of state, $\omega(t)$ at 3.5 sigma level. It is against the constant equation of state, $\omega = -1$ for the cosmological constant. The 3D map of galaxies to be initiated by Dark Energy Spectroscopic Instrument (DESI) (2018 onward) is expected to improve the observations.

Second, measurements of the Hubble parameter in the local universe based on gravitational lensing, performed by the HoLiCOW collaboration (Bonvin et al. 2017) have shown larger value of the parameter when compared with the most satisfactory result of the Planck satellite coming from analysis of the cosmic background radiation. Time delay of light rays of the secondary images of high z quasars which are lensed by the massive lens, HE 0435 – 1223 has shown $H_0 = 71.9 \pm 2.7$ km/sec/Mpc which is not congruent with the Planck data (Ade et al. 2016), $H_0 = 67.9 \pm 1.5$ km/sec/Mpc. A time varying equation of state with appropriate parameterisation shows natural tendency for an increased expansion rate. Dynamical scalar fields also belong to theoretical possibilities.

Prospect of discovery of dark energy and the recent anomalies, is now to precisely determine the expansion history of the universe and constrain the equation of state at various epochs. Some of the dynamical fields are associated with deep mysteries of fundamental physics such as variation of the fundamental constants.

### 5.7 Pulsars, quasars and the fundamental constants

The spacetime description (metric) of gravity appears naturally out of universality of gravity (principle of equivalence) which demands non-existence of non-universal fields such as the scalar degrees of freedom that occur in scalar – tensor theories (Jordan – Brans – Dicke theory and its generalisations). Presence of non-universal fields in addition to the metric is associated with variation of the Cavendish constant (G) which is prohibited by the strong principle of equivalence. This principle also restricts variation of non-gravitational constants such as the electromagnetic fine structure constant ($\alpha_e = e^2 / 4\pi\varepsilon_0 \hbar c \approx 1/137$). As these are physical principles they are testable.

Variation of $\alpha_e$ has been subject to astronomical tests through observation of absorbing intergalactic gases in the line of sight to distant quasars. If $\alpha_e$ were different in the remote (early) universe, it would change the energy levels of atoms in the gas clouds which in turn would affect the spacing of the absorption spectral lines. A 2004 study through the VLT showed that over the last billion year of cosmic history, $\alpha_e$ has changed by an amount $\delta\alpha_e / \alpha_e \approx 6 \times 10^{-7}$ (Quast, Reimers and Levshakov 2004). But the case is not closed. In 2011, a combined study of VLT and Keck telescopes has shown that the "constant" can have spatial variation as large as $10^{-5}$ (Webb et al. 2011). Large number of high resolution observations of the quasar absorption systems will be necessary to draw inference on this issue.

Timing analysis of a pulsar – white dwarf binary has, on the other hand, shown that G has negligible position variation (Zhu et al. 2015). Future pulsar timing arrays to be studied by the upcoming radio telescope, SKA will shed more light on this matter.

Variation of fundamental constants, theory of gravity and therefore, dark matter and dark energy are inextricably connected to each other. A congruent description is not yet in sight. It is a challenge for the future generation of large telescopes.

## 6. Conclusion

From its majestic mathematical ivory tower, GR has begun its true scientific journey only recently, thanks to the runaway progress in theory and observations. Understanding spacetime reveals new windows to unravel mysteries of the astrophysical world, be it spin of the black holes, gravitational waves or the dark universe. The elegant mathematics of differential geometry has given us a theory which has explained wide variety of phenomena in the universe and has generated new and challenging problems in the natural science of astronomy. We still do not know why the theory, formulated without experimental need and purely based on elegant physical principles, is so precise and at the same time gives way to unexpected avenues that may be lurking beneath the theory itself. The theory also hosts certain physical principles required for a fundamental theory of spacetime which may ignite several observational tests in near future with accuracies previously unanticipated. The challenge today is that there are more secrets hidden in the theory and much work is left to be done. Relativistic astrophysics is now the frontier in understanding the universe. In the coming years, the theory will undergo observational tests not only in tiny regions of spacetime but also in extremely large scale of cosmic time spanning almost more than 50% of the age of the universe. The challenges are going to broaden the existing horizons by the time we enter the coming century.

.